\documentclass[iop,apj,tighten]{emulateapj}

\usepackage{apjfonts}
\usepackage{amsmath,amstext}
\usepackage{verbatim}
\usepackage[breaklinks,colorlinks,citecolor=blue,urlcolor=magenta,linkcolor=blue]{hyperref}
\usepackage[toc,page]{appendix}
\usepackage{ulem}
\usepackage{bm}
\usepackage{graphicx}
\usepackage{multimedia}
\usepackage{float}
\usepackage[utf8]{inputenc}
\usepackage{newunicodechar}
\usepackage{enumerate}
\newcommand{\textprime}{\ensuremath{'}}
\newunicodechar{′}{\textprime}

\defcitealias{YMJ:2017}{YMJ17}
\defcitealias{Gordon:2016}{G16}
\defcitealias{Gordon:2003}{G03}
\defcitealias{Weisz:2013mc}{W13}
\defcitealias{WD:2001}{WD01}
\defcitealias{Dolphin:2002}{D02}
\defcitealias{Fitzpatrick:1999}{F99}
\defcitealias{DeMarchi:2014fe}{DM14a}
\defcitealias{DeMarchi:2014en}{DM14b}
\defcitealias{DeMarchi:2016}{DM16}

\newcommand{\angstrom }{\mbox{\normalfont\AA}}

\newcommand{\about}{\textasciitilde}
\newcommand{\fsf}{\textit{F475W}}
\newcommand{\eof}{\textit{F814W}}

\shorttitle{SMIDGE SMC 3D Geometry and Dust Extinction with Red Clump \& Red Giant Branch Stars}
\shortauthors{Yanchulova Merica-Jones et al.}

\begin{document}

\title{Three-Dimensional Structure and Dust Extinction in the Small Magellanic Cloud}
\author{Petia Yanchulova Merica-Jones\altaffilmark{1}, Karin M. Sandstrom\altaffilmark{1}, L. Clifton Johnson\altaffilmark{2}, Andrew E. Dolphin\altaffilmark{3}, Julianne J. Dalcanton\altaffilmark{4}, Karl Gordon\altaffilmark{5, 6}, Julia Roman-Duval\altaffilmark{5}, Daniel R. Weisz\altaffilmark{7}, Benjamin F. Williams\altaffilmark{4}}
\affil{$^1$Center for Astrophysics and Space Sciences, Department of Physics, University of California, 9500 Gilman Drive, La Jolla, San Diego, CA 92093, USA}
\affil{$^2$Department of Physics and Astronomy, Northwestern University, 2145 Sheridan Road, Evanston, IL 60208, USA}
\affil{$^3$Raytheon; 1151 E. Hermans Road, Tucson, AZ 85756, USA}
\affil{$^4$Department of Astronomy, University of Washington, Box 351580, Seattle, WA 98195, USA}
\affil{$^5$Space Telescope Science Institute, 3700 San Martin Drive, Baltimore, MD 21218, USA}
\affil{$^6$Sterrenkundig Observatorium, Universiteit Gent, Gent, Belgium}
\affil{$^7$Department of Astronomy, University of California, 501 Campbell Hall \#3411, Berkeley, CA 94720-3411, USA}

\begin{abstract}
We examine the three-dimensional structure and dust extinction properties in a \about 200 pc $\times$ 100 pc region in the southwest bar of the Small Magellanic Cloud (SMC).  We model a deep Hubble Space Telescope optical color-magnitude diagram (CMD) of red clump and red giant branch stars to infer the dust extinction and galactic structure.  We model the distance distribution of the stellar component with a Gaussian and find a centroid distance of 65.2 kpc (distance modulus $\mu$ = 19.07 mag) with a FWHM $\approx$ 11.3 kpc.  This large extent along the line of sight reproduces results from previous studies using variable stars and red clump stars.  Additionally, we find an offset between the stellar and dust distributions, with the dust on the near side relative to the stars by 3.22 $^{+1.69}_{-1.44}$ kpc, resulting in a 73\% reddened fraction of stars.  Modeling the dust layer with a log-normal $A_V$ distribution indicates a mean extinction $\langle A_V \rangle$ = 0.41 $\pm$ 0.09 mag.  We also calculate $A_V/N_H$ = 3.2 - 4.2 $\times10^{-23}$ mag cm$^2$ H$^{-1}$ which is significantly lower than the Milky Way value but is comparable to previous SMC dust-to-gas ratio measurements.  Our results yield the first joint dust extinction and 3D geometry properties in a key region in the SMC.  This study demonstrates that CMD modeling can be a powerful tool to simultaneously constrain dust extinction and geometry properties in nearby galaxies.
\end{abstract}



\section{Introduction} \label{sec:intro}

Understanding dust and dust extinction at low metallicity is important for a number of reasons.  Dust in galaxies attenuates starlight, and affects the star formation, radiative transfer, thermodynamics, and chemistry in the interstellar medium (ISM).  The amount and type of dust is likely strongly dependent on metallicity \citep[e.g.][]{1998:Dwek, 1999:Zubko, 2003:Clayton, 2006Ulysses, 2008:Zhukovska, 2013Asano, 2014:Remy-Ruyer, 2014:Roman-Duval, 2015Feldmann, 2019Chastenet, 2019DeVis}, making low-metallicity dust studies critical due to the prevalence of these conditions in the early universe.

Fundamental dust extinction properties in low-metallicity environments show distinct differences from the Milky Way.  For example, the observed extinction curve of the Small Magellanic Cloud (SMC) at 1/5 - 1/8 $Z_{\odot}$ \citep{Dufour:1984,RussellDopita:1992,Kurt:1999,Lee:2005,Rolleston:1999,Rolleston:2003} shows a steep UV rise and a weak or absent 2175 $\angstrom$ bump \citep{Lequeux:1982, Prevot:1984, Gordon:1998, Gordon:2003, MaizApellaniz:2012}.  These differences may hold clues to how the dust in such environments behaves and evolves, in turn affecting galaxy evolution.

Accounting for dust extinction at high-redshift can be challenging \citep{2001:Somerville}, but fortunately a great deal can be learned from nearby low-metallicity galaxies which can be considered analogues to the high-redshift universe where galaxies must have formed at a very low metallicity \citep{Madau:2014}.  The SMC in particular is an especially suitable target due to its proximity, with its center at $\sim$ 62 kpc \citep{degrijs:2015, Scowcroft:2016hm}.  Indeed, "SMC-like" extinction is the standard used to account for dust at high-redshift and/or low metallicity \citep{Gordon:2003}, thus it is a natural target for low-Z studies of dust properties.  

The SMC's proximity also allows us to study in detail its geometry, including the distance to the stars and the ISM, and the relative positions of these two galactic components.  One of the goals of our study, named the Small Magellanic Cloud Investigation of Dust and Gas Evolution (SMIDGE) \citep[][hereafter YMJ17]{YMJ:2017}, is to examine dust extinction and 3D geometry holistically and derive resolved maps of extinction ($A_V$ and $R_V$) for low metallicity dust.

Traditionally, the shape of dust extinction curves are studied using the pair method \citep{Trumpler:1930, Massa:1983, Cardelli:1992} which compares the spectra of a reddened and an unreddened star of approximately the same spectral type.  However, this method relies on individual lines of sight to a select sample of UV-bright stars with high signal-to-noise data.  An efficient and complementary technique to measure dust extinction and the extinction curve is to use multiband photometric observations of resolved stars to measure the average extinction properties in a galactic region.  Using resolved stellar photometry to measure dust extinction, however, may require modeling the galactic geometry as well.

\begin{figure*} 
  \centering 
    \includegraphics[width=1.0\textwidth]{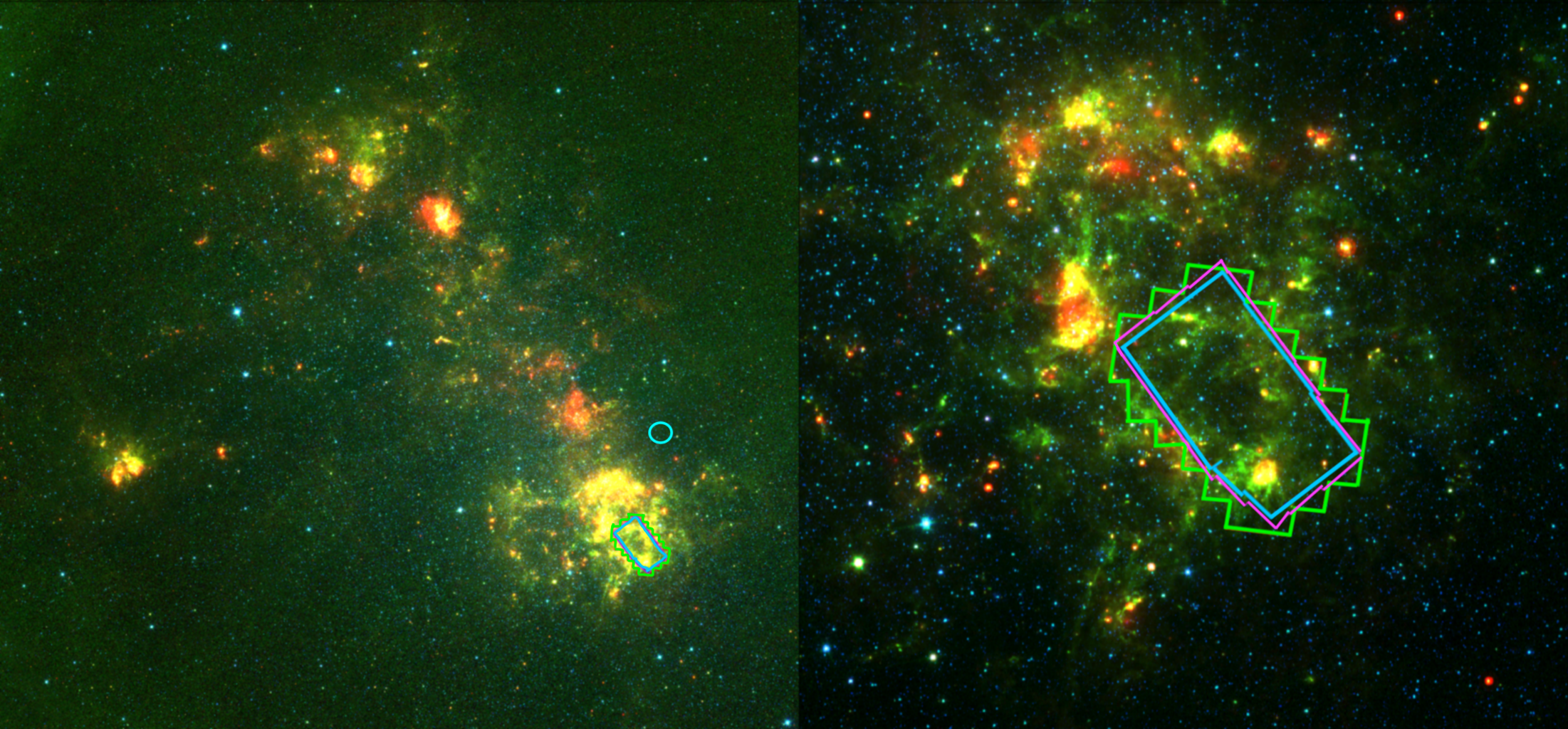}
  \caption{The SMIDGE survey field (100 $\times$ 200 pc) in the SW bar of the SMC is overlaid on a Spitzer Space Telescope composite image in 3.6, 8, and 24 $\mu$m from the SAGE-SMC survey \citep{Gordon:2011jq}.  The \textit{HST} imaging footprint is shown for ACS/WFC (green), WFC3/UVIS (magenta), and WFC3/IR (blue). The cyan circle indicates the \citet{Weisz:2013mc} SMC-2 field used in our star formation history analysis.}
  \label{fig:smidge_spitzer_sage}
\end{figure*}

Interstellar dust is expected to affect all color-magnitude diagram (CMD) features.  These effects depend on its distribution relative to the stars.  If the dust is in a foreground screen relative to the stars, the stars will be reddened to various degrees depending on the ISM column density distribution, which is often observed to be log-normal \citep{Hill:2008, Kainulainen:2009, Hennebelle:2012}.  If it is mixed with the stars, some stars will experience only a fraction of the overall reddening.  If the dust is in a thin layer and the stars are in a spatially extended distribution (assuming they are not embedded in the dust layer), a part of the CMD will be foreground to the dust and unreddened, while the rest will experience the full reddening from the dust layer.  In general, CMD features that are compact or perpendicular to the reddening vector are the most useful for measuring the effects of dust and geometry.  CMD modelling of these effects can give clues about the relative line of sight depths of the dust and the stars and the reddened fraction of stars.  The models can then be used to produce a map of the extinction and/or the geometry.  Extinction mapping of reddened stars using color-color or color-magnitude diagrams has been the subject of numerous Milky Way \citep{2001Lombardi, Nataf:2013, Schlafly:2016}, M31 \citep{Dalcanton:2015bl}, LMC \citep{2007Imara, 2008Dobashi, Choi2018}, and SMC studies (\citealt{Gardiner:1991, 2009:Dobashi}, \citetalias{YMJ:2017}).

Geometric effects, such as an elongation of the galaxy along the line of sight spreads the stars in magnitude, but not color.  Dust, on the other hand, affects both the color and magnitude of stars in a CMD.  Features that would appear narrow in the absence of dust tend to get spread out when there is dispersion in the amount of extinction due to the column density structure of the ISM.  If this effect can be modelled, then we can measure the mean and the width of the extinction distribution.  By finding the stars' observed displacement in a CMD from a predicted theoretical unreddened location, we can calculate the dust extinction in magnitudes, $A_{\lambda}$, where $\lambda$ is the effective wavelength of each photometric band.  We can measure $R_\lambda=A_\lambda/(A_\lambda\prime-A_\lambda\prime\prime)$ (related to the dust grain size distribution) from a single CMD color.  Measuring $R_\lambda$ across photometric bands from IR to UV allows us to sample the extinction curve shape.

The SMC, however, presents a unique challenge due to the combined effects of its extinction, relative proximity and its appreciable elongation along the line of sight \citep{Subramanian:2009,Haschke:2012smc,JD:2016a}.  For example, in \citetalias{YMJ:2017} we found that to accurately measure the SMC extinction curve from the slope of reddened red clump (RC) stars, we also needed to account for the galactic depth along the line of sight.  This effect, due to the appreciable spread in stellar distances, could give the impression of a steeper RC slope since stars behind the dust layer will have fainter magnitudes at increasing distances.  This observation would in turn seem to imply higher $R_\lambda$ values.  We accounted for the depth by employing a simple model for the RC which specified a theoretical unreddened RC location based on the local star formation history (SFH), and a set of parameters for the extinction distribution, the extinction law, and the galactic depth along the line of sight.  

In this paper, we build on our previous work and use the full RC and red giant branch (RGB) CMD to model the dust extinction, 3D geometry, and the SFH holistically.  The effects of dust are ubiquitous in the SMIDGE data, and beside spreading the reddened RC into a streak, dust also widens the reddened RGB.  The signature of the SMC's geometry is most clearly seen in the double RGB which is a result of the relative thicknesses along the line of sight and the absolute positions of the dust and stellar components of the galaxy.  The SMC's depth is not as immediately obvious from the CMD, but it nonetheless results in a measurable and significant spread in the magnitude of both the RC and the RGB.  Analyzing the RC and the RGB simultaneously, versus only the RC, allows us to better constrain the CMD stellar number counts statistically (and thus the SFH), and to separately account for the effects of galactic geometry and dust extinction.  The reason for this is that the RC reddening vector slope is a product of both the extinction curve and galactic geometry (where in \citetalias{YMJ:2017} we showed these can be degenerate without additional information).  At the same time, while it would be challenging to constrain the slope of the extinction vector using the RGB alone, the RGB very strongly constrains the relative offset between the dust and the centroid of the stellar distribution.  Modeling the combined effects of the RC and the RGB in theory enables one to separate these effects and better constrain the slope of the extinction vector.

Accurately modeling the CMD requires a representative SFH as a starting point.  There have been numerous efforts to obtain the SFHs in the SMC by modeling CMDs of resolved stellar populations \citep{Tosi:1989, 1996TolstoySaha, HarrisZaritsky:2001, Dolphin:2002, 2002:Williams, ZaritskyHarris:2004, Cignoni:2009, Weisz:2013mc, Rubele:2015vmc, 2017:Williams, Rubele:2018}.  These models commonly need to account for the initial mass function (IMF), binary fraction, stellar evolution, time resolution, metallicity, distance, extinction, and photometric noise.  Typically, the distance is a parameter specified by ancillary studies.  Although some approaches can also account for the extinction within a galaxy, to simplify the solution most studies preferably examine regions with low internal $A_V$ which is treated in a minimal way.  Deriving the SFH itself seems to be unaffected by the presence of a significant galactic depth \citep{HarrisZaritsky:2004, Rubele:2018}.

In this paper, we forward model the SMIDGE RC and RGB CMD with realistic noise and a model which includes geometry, distance, and dust.  We first create synthetic model CMDs based on a set of input SFHs and stellar evolution parameters.  Then we model the line-of-sight depth of the stellar distribution, the relative offset of the dust with respect to the stars, and the dust extinction distribution.  Using a realistic model of photometric uncertainties, we compare model CMDs to SMIDGE observations to measure properties of the SMC's 3D distribution of gas and dust.

We briefly describe the SMIDGE data in Section \ref{sec:obs}.  We discuss the CMD modeling process in Section \ref{sec:cmdmatching}, including details about how we {match CMDs and} account for the uncertainties.  Our results are presented in Section \ref{sec:results} and discussed in Section \ref{sec:discussion}.  We conclude with Section \ref{sec:conclusions} and provide more relevant information in the Appendix.

\begin{figure} 
   \centering
   \includegraphics[width=0.48\textwidth]{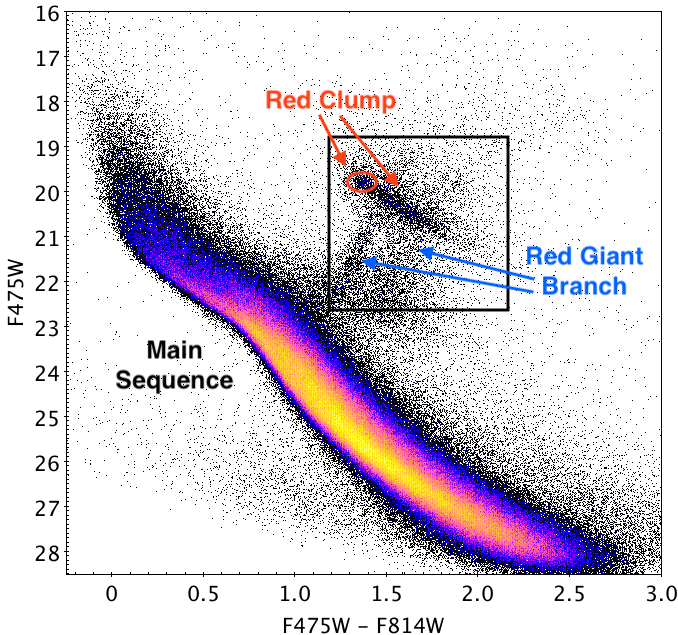}
   \caption{The SMIDGE $\fsf-\eof$ optical color-magnitude shows the boundaries of the red clump and red giant branch region analyzed here.  Dust extinction is evident from the extended reddened RC (which would theoretically be located within the red ellipse in the absence of dust) and the extended and bimodal reddened RGB which would be represented by a single vertical sequence if unreddened).  The observed bimodal RGB indicates that the dust layer is thin relative to the stellar distribution, with few embedded stars.  These combined effects cause a distinct unreddened and reddened stellar populations.}
   \label{fig:smidgeCMD}
\end{figure}

\section{Data}\label{sec:obs}
The SMIDGE survey (GO-13659) consists of eight-band \textit{Hubble Space Telescope} (HST) imaging of a \about 100 pc $\times$ 200 pc region in the SMC southwest bar.  Its location is shown in Figure \ref{fig:smidge_spitzer_sage}. We refer to \citetalias{YMJ:2017} for the details of the SMIDGE imaging footprint, HST camera and filter selections, photometry processing, and data culling.  Also see the SMIDGE survey paper (Sandstrom et al., in preparation) for complete details.  This work is based on the $\fsf-\eof$ vs $\fsf$ CMD shown in Figure \ref{fig:smidgeCMD} (where the units of all magnitudes are Vega magnitudes).

\section{Modeling the SMIDGE Color-Magnitude Diagram} \label{sec:cmdmatching}


Our goal is to constrain the SMC dust extinction and 3D geometry parameters by finding the best matching model CMDs to the SMIDGE observations.  The best match is determined by comparing the number of stars in bins of color and magnitude between the modeled and the observed CMD. The optical CMD is ideal for this purpose due to its high signal-to-noise (compared to other available filter combinations), and broad wavelength baseline.

The location and number of stars in the CMD are mainly determined by these factors:

\begin{enumerate}
  \item The SFH, i.e. the star formation rate SFR($t$) and metallicity evolution Z($t$), which defines the density, unreddened position and morphology of the stellar populations,
  \item The distribution of dust extinction, which sets the additional spread in color and magnitude of the stars,
  \item The extinction law, which sets the slope of the reddening vector,
  \item The average distance modulus which determines the overall apparent magnitude of the stars,
  \item The line-of-sight depth, which affects the spread in magnitude of the stellar populations, and,
  \item The dust-stellar centroid offset (which we simply call the dust-stars offset), which, along with an assumption of the relative thickness of the dust and stellar distributions, determines the fraction of reddened stars and their distance distribution.
\end{enumerate}

To model these properties, we take a forward modeling approach to CMD matching by creating synthetic CMDs which include the effects of the SFH, dust extinction, galactic geometry, and observational noise, and compare the number counts in the CMD to find the best match.  Versions of CMD matching have been used by several studies (see references in Sec. \ref{sec:intro}).  The method we use is based on the approach presented in \citet[][hereafter D02]{Dolphin:2002} who developed a widely used technique for CMD modelling \citep[e.g.][]{Dolphin:1997, Aparicio:1997} to derive the SFHs of nearby galaxies with spatially resolved stellar populations \citep{2009TolstoyHillTosi, Weisz:2013mc, 2014:Weisz}.  \citetalias{Dolphin:2002} created a maximum likelihood CMD fitting package -- \texttt{MATCH} -- to recover the best SFH by modelling the observed stellar density in a Hess diagram (a binned CMD showing the relative density of stars).  We rely on the \citetalias{Dolphin:2002} approach conceptually to search combinations of dust extinction and geometry parameters for the best-fit CMD.  

Although \texttt{MATCH} can also model a dust extinction distribution, it is not designed to model extinction and 3-D geometry (a spread of stellar distances and a dust-stars offset) simultaneously.  Since the SMC geometry necessitates that we take into account all of these features at the same time, we do not use \texttt{MATCH} per se to find the SMIDGE SFH or to fit the SMIDGE CMDs, but only use it to generate single-distance, unreddened synthetic CMDs.  We build on the \texttt{MATCH} foundation by developing a fitting method which facilitates our use of more complex models for distance and extinction distributions with the \texttt{syncmd} \footnote{\texttt{syncmd} as of 2020 February 29 (C. Johnson): \url{https://github.com/lcjohnso/syncmd/tree/simple_attenuation}} code.  \texttt{syncmd} applies distance, reddening effects, and a noise model using functions available from the Bayesian Extinction and Stellar Tool (BEAST) \footnote{\texttt{BEAST} as of 2020 September 25: \url{https://github.com/BEAST-Fitting/beast}} of \citet[][hereafter G16]{Gordon:2016} to create CMDs which we can directly compare to the data.

\begin{figure*} 
  \centering 
    \includegraphics[width=0.49\textwidth]{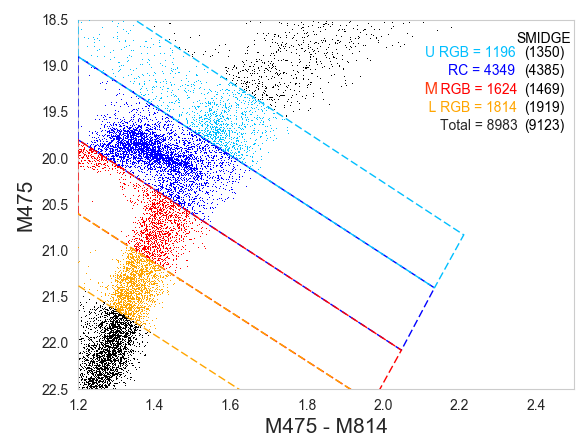}
    \includegraphics[width=0.49\textwidth]{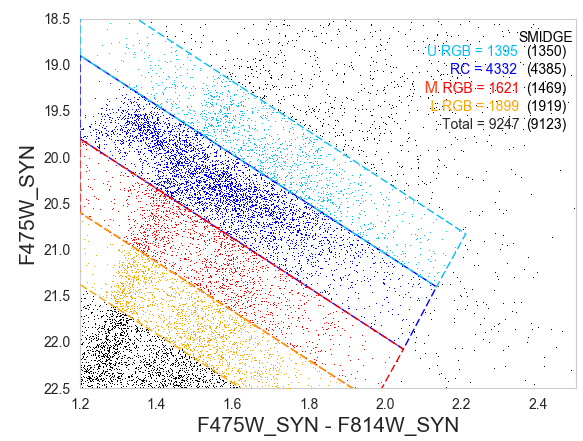}
   \caption{(Left) The unreddened synthetic CMD generated by \texttt{MATCH/fake} \citep{Dolphin:2002} based on the best-fit SMIDGE SFH chosen as described in Section \ref{sec:modelingSFH}.  (Right) The same synthetic CMD processed with observational noise, dust extinction and geometry effects ( Sections \ref{sec:addGeometry} \ref{sec:addExtinction}).  The CMD is divided in the following subregions marked by diagonal bins: upper RGB (URGB; light blue), RC \& RGB (blue), mid RGB (MRGB; red), and low RGB (LRGB; orange).  A comparison of stellar number counts (in both the unreddened and the reddened CMDs) is made with the same regions in the SMIDGE field.  The diagonal orientation of the bins follows the slope of the reddening vector (subject to SMC Bar extinction \citep{Gordon:2003}, as found in \citetalias{YMJ:2017}).  The boundaries of the colored boxes coincide with the boundaries of the RC/RGB region chosen here and in the rest of our analysis.  The stellar number counts are shown in the legend and are color coded correspondingly.  The respective SMIDGE stellar number counts for each region are in black.  In the reddened CMD on the right stars in CMD features outside of the defined RC/RGB region are reddened into some of the subregions and increase the number of stars there. Other regions (the RC) experience a negligible decrease in the number of stars due to a low/no contribution from CMD features lying in the opposite direction of the reddening vector and due to possible displacement outside of the defined reddening vector boundaries.}
   \label{fig:syncmdregions}
\end{figure*}

In principle, finding the best fit model could involve searching through a large grid of CMD models covering all possible variables.  The dimensions of this grid would be the SFH($t_{i}$) and Z($t_{i}$) for time bins $t_{i}$; the dust extinction parameters; and the 3D geometry parameters.  Depending on the number of time bins in the SFH, such a calculation can become very computationally expensive.  We therefore simplify the procedure by making these choices:

\begin{enumerate}
  \item We focus only on the RC/RGB region
  
These populations are intrinsically narrow and most strongly display the effects of dust extinction and 3D geometry as seen in Figure \ref{fig:smidgeCMD}, where the RC is extended and the RGB is extended and doubled.  We thus draw a selection box around this region.  To model the dust and geometry we do not need to include the whole RGB, nor do we have to cut the selection box at fainter magnitudes with an exquisite precision.  On the other hand, if the selection box is extended to fainter magnitudes, we would end up selecting main sequence stars which is not optimal.  Therefore, the results should not be affected by expanding the box.  We define the RC/RGB region as stars with $\fsf-\eof$ color of 1.2 - 2.2 mag and $\fsf$ magnitude of 18.5 - 22.75 mag and obtain a CMD subset of $\sim$12,000 stars. These ranges encompass the unreddened RC, the streak of the reddened RC, and the tip and base of the RGB.

  \item We simplify the SFH since we only model the RC/RGB

Since finding the SFH is not our main goal, and the extinction and geometry parameter fitting only requires a broad knowledge of the SFH, the RC/RGB SFH can be treated coarsely with limited time bins.  See Section \ref{sec:modelingSFH} for details.

  \item We separate the SFH and the dust/geometry grid search
  
We first measure the SFH by matching the stellar number counts in coarse RC/RGB bins (see Fig. \ref{fig:syncmdregions}).  After determining the approximate SFH, we proceed to search the full grid of dust and geometry parameters.
\end{enumerate}

Using the choices above, the conceptual outline of the CMD matching process is the following, with details outlined in the rest of Section \ref{sec:cmdmatching}.

\begin{enumerate}[I.]
  \item Fit a star formation history for the SMIDGE field. (Section \ref{sec:modelingSFH}).
  \begin{enumerate}
     \item First generate unreddened, single-distance synthetic CMDs according to a range of star formation rates and small metallicity offsets based on adjustments of a known SFH of a nearby region.
     
     Compare RC/RGB stellar number counts between the unreddened synthetic and the observed CMDs in coarse bins designed to be insensitive to extinction and geometry (left panel of Fig. \ref{fig:syncmdregions}). Select SFHs producing CMD statistics consistent with the SMIDGE stellar number counts. (Sec. \ref{sec:unreddenedCMD}.)
     \item Refine the SFH selection by applying a fixed set of geometry, distance, and dust extinction parameters (obtained in YMJ17) to the CMDs above.  Model the SMIDGE distance.  Also include realistic observational noise (see Part II).  Recalculate the stellar number counts in the RC/RGB coarse bins. Perform a second SFH refinement using a narrower range of the cuts. (Sec. \ref{sec:addDistance} - \ref{sec:addNoise}.)
     \item Select the SFH producing a best-fit coarse CMD and use it as a foundation for the analysis in Part \ref{itm:part2}.
  \end{enumerate}
  \item Fit the dust extinction and geometry model combinations. (Section \ref{sec:addExtinction} - \ref{sec:uncertainty}). \label{itm:part2}
  \begin{enumerate}
      \item Generate model CMDs, with the SFH from step I, in a grid of combinations of dust extinction and 3D galactic geometry parameters.
      \item Add observational noise to the stars in the model CMDs as characterized using artificial star tests.
      \item Find the best fitting model and the uncertainty in the results by measuring the probability density function of the model grid.
  \end{enumerate}
\end{enumerate}

\begin{deluxetable}{ccc} 
\tabletypesize {\small}
\tablecolumns{3}
\tablecaption{Grid of SFH parameters}
\tablehead{ \colhead{Time [Gyr ago]} & \colhead{SFR [$10^{-4}M_{\odot}/yr$]} & \colhead{[M/H]}}
\startdata
 0.004 -- 0.01 & 1.0 -- 3.0 & -0.65  --  -0.45\\
 0.01 -- 0.03 & 1.0 -- 3.0 & -0.65  --  -0.45\\
 0.03 -- 0.05 & 1.0 -- 3.0  & -0.65  --  -0.45\\
 0.05 -- 0.16 & 1.0 -- 3.0  & -0.65  --  -0.45\\
 0.16 -- 0.25 & 1.0 -- 3.0  & -0.65  --  -0.45\\
 0.25 -- 0.40 & 1.0 -- 3.0  & -0.65  --  -0.45\\
 0.40 -- 0.63 & 6.0 -- 9.0 & -0.70  --  -0.60\\
 0.63 -- 1.0 & 1.5 -- 4.0 & -0.75  --  -0.65\\
 1.0 -- 1.26 & 1.5 -- 4.0 & -0.75  --  -0.65\\
 1.26 -- 1.59 & 1.5 -- 4.0 & -0.75  --  -0.65\\
 1.59 -- 2.5 & 5.0 -- 8.0  & -0.95  --  -0.85\\
 2.5 -- 3.2 & 5.0 -- 8.0  & -1.10  --  -1.0\\
 3.2 -- 4.0 & 1.5 -- 3.0  & -1.10  --  -1.0\\
 4.0 -- 6.3 & 1.5 -- 3.0  & -1.20\\
 6.3 -- 10.0 & 1.5 -- 3.0  & -1.40\\
 10.0 -- 14.13 & 1.5 -- 3.0  & -1.70\\
 \enddata
\tablecomments{SFR and metallicity values in each age bin are equally spaced within the range given: 4 - 400 Myr age bins assume 5 values; 400 Myr - 3.2 Gyr assume 4 values, and bins with ages $\geq$ 3.2 Gyr assume 3 values. The metallicity dispersion at each time bin is 0.25.}
 \label{tab:sfhvary}
\end{deluxetable}

\begin{deluxetable}{lccc}
\tabletypesize {\footnotesize} 
\tablecolumns{5}
\tablewidth{0.45\textwidth}
\tablecaption{RC/RGB region stellar number counts for SFH selection}
\tablehead{ \multicolumn{1}{l}{Region} & \colhead{All SFH Models} & \colhead{SMIDGE} & \colhead{Best-fit SFH}}
\startdata
 Red Clump & 3000-6000 & 4385 & 4332\\
 Upper RGB & 800-1600 & 1350 & 1395\\
 Mid RGB & 1000-2500 & 1469 & 1621\\
 Low RGB & 1100-2400 & 1919 & 1899\\
 Total RC + RGB & 6000-12000 & 9123 & 9247\\
 \enddata
\tablecomments{The Region column corresponds to each of the RC/RGB regions in Figure \ref{fig:syncmdregions}. The ``All SFH Models'' column shows the full range of stellar number counts in each region sampled by the combinations of SFH parameters in Table \ref{tab:sfhvary}.  The SFH cuts we impose are based on the SMIDGE stellar number counts in a region $\pm$ a percentage of the full range of this same region for all SFH models. The cuts are, (a) $\pm$ 10 \% for the initial unreddened CMD models, and (b) $\pm$ 5\% for the final reddened CMD models (there is a small deviation from these cuts for the mid RGB region (red in Fig. \ref{fig:sfhpanels}) due to a horizontal branch contribution noted towards the end of Section \ref{sec:modelingSFH}). }
 \label{tab:sfhstats}
\end{deluxetable}


\begin{figure} 
   \centering
   \includegraphics[width=0.47\textwidth]{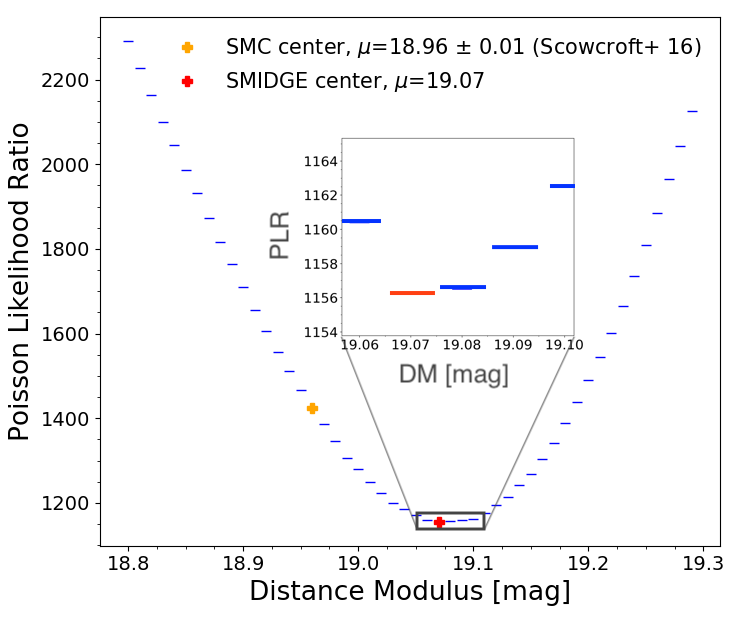}
   \caption{ Distance modulus test for the SMIDGE field using a Poisson likelihood ratio test (see Section \ref{sec:pdfanalysis}).  The SMC center is measured to be at $\mu$=18.96 $\pm$ 0.01 mag by \citet{degrijs:2015} and \citet{Scowcroft:2016hm}, while we measure the center of the SMIDGE field to be located at $\mu$=19.07 mag.  The greater SMIDGE distance is expected from models and measurements of the extended stellar distribution.}
   \label{fig:dmodtest}
\end{figure}

\begin{figure} 
  \centering
    \includegraphics[width=0.47\textwidth]{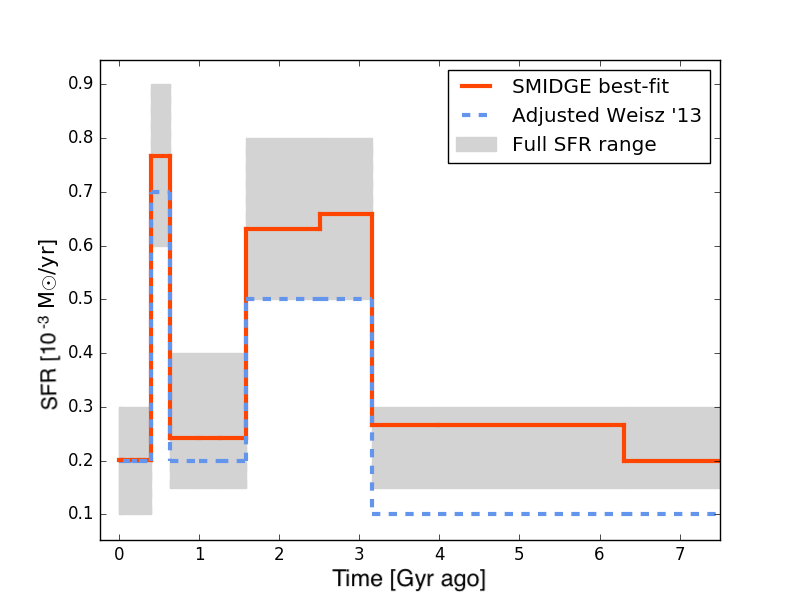}
   \caption{The SMIDGE SFR derived as described in Section \ref{sec:modelingSFH}.  Adjustments to the \citetalias{Weisz:2013mc} SFR are shown in dashed blue and were used in our \citetalias{YMJ:2017} work.  We allow the SFR to vary within the range in the gray band with the resolution specified in Table \ref{tab:sfhvary}.  The SFR remains constant for ages beyond 7 Gyr.  Our results for SMIDGE are in red and show SFR enhancements to the \citetalias{Weisz:2013mc} SFR at almost all ages, including at times where our SFR is allowed to vary below their result.  We see a SFR at older ages ($\geq$ 3 Gyr) which is enhanced by a factor of 2.}
   \label{fig:sfhpanels}
\end{figure}

\subsection{Forward Modeling the CMD} \label{sec:forwardModel} 

\subsubsection{Unreddened Single-Distance CMD} \label{sec:unreddenedCMD}

We can predict the number density of a population of stars in the CMD given a SFR($t$) and a Z($t$).  We use \texttt{MATCH}'s \texttt{fake} CMD simulation tool \citepalias{Dolphin:2002} to generate synthetic CMD models without noise, extinction or geometric effects such as the one in the left panel of Figure \ref{fig:syncmdregions}.  First we create a synthetic CMD based on adjustments to the \citet[][hereafter W13]{Weisz:2013mc} SMC SFH for a nearby region.  Our motivation for starting with this SFH is detailed in Sec. \ref{sec:modelingSFH} where we discuss other details of the SFH analysis.  We input the following fixed parameters into \texttt{fake}: a \cite{Kroupa:2001} initial mass function, PARSEC stellar evolution models \citep{Bressan:2012, Tang:2014, Chen:2015}, a binary fraction of 0.35, a foreground MW extinction of $A_V$ = 0.18 mag derived from MW {\sc HI} foreground towards the SMIDGE field \citep{Muller:2003, Welty:2012}, and a distance modulus $\mu$ = 18.96 mag for the center of the SMC \citep{degrijs:2015, Scowcroft:2016hm}.

\subsubsection{Modeling the SMIDGE Distance} \label{sec:addDistance}

To properly model the CMD, we must also know the average distance to the stars in the region.  The SMC's mean distance modulus varies across the galaxy due to SMC's three-dimensional structure.   \citet{Scowcroft:2016hm} measure a distance modulus of $\mu$ = 18.96 $\pm$ 0.01 mag for the center of the SMC using Cepheid variables, and \citet{degrijs:2015} also find a mean SMC $\mu$ = 18.96 $\pm$ 0.02 mag.  \citet{Haschke:2012smc} find a distance modulus varying between 18.94 mag $\leq \mu \leq$ 19.17 mag using RR Lyrae stars and Cepheids in the bar and wing of the SMC.  \citet{Nidever:2013} use RC stars to measure the SMC line-of-sight depth and find $\sim$23 kpc for the eastern side and $\sim$10 kpc for the western side.  \citet{Ripepi:2017} use classical Cepheids and find a distance spread in the SMC SW Bar of 62.5 - 65 kpc ( 18.98 $\leq \mu \leq$ 19.06 mag). Models of the SMC and the LMC \citep{Besla:2007, Besla:2012, Besla:2016}, show that the Magellanic Clouds have experienced repeated interactions with each other which cause asymmetry in the stellar structures of the galaxies.   Additionally, it is well-known that the SMC has a large depth along the line of sight \citep{Subramanian:2009, Subramanian:2012, Subramanian:2017, JD:2016a, JD:2016b}.  

We investigate the distance modulus appropriate for the SMIDGE field in the SW Bar by performing a preliminary Poisson likelihood ratio CMD matching calculation using our previous knowledge of the SFH, extinction, and geometry we obtained in \citetalias{YMJ:2017}, namely: an extinction law consistent with SMC Bar \citet{Gordon:2003} extinction, a log-normal median $\widetilde{A_V}$ = 0.32 and width $\langle{\sigma_{A_{V}}}\rangle$ = 0.3, a dust distance $D_{dust}$ = 60 kpc (dust-stars offset of $\sim$ 2 kpc), and a 10-kpc line-of-sight depth (resulting in a 0.65 fraction of reddened stars).  Using the adjusted \citetalias{Weisz:2013mc} SMC SFH, we proceed to shift the CMD in distance.  The dust extinction and 3D geometry parameters are held constant as the synthetic CMDs are shifted in distance (magnitude) over the range 18.8 mag $\leq \mu \leq$ 19.3 mag.  Since the distance calculation relies only on shifting the CMD in magnitude, using a preliminary version of these parameters does not affect the results in terms of their sensitivity to a particular set of parameter values.  Comparing to the SMIDGE observations, we find a best match at $\mu$=19.07 mag as shown in Figure \ref{fig:dmodtest}.  The larger distance is expected for the SMIDGE region based on models of the extended stellar distribution placing SMC's NE regions closer than the SW regions \citep[i.e.][]{Haschke:2012smc}.

\subsubsection{Modeling the 3D Geometry} \label{sec:addGeometry}

We consider two aspects of the three-dimensional geometry which impact the SMIDGE CMD: the depth along the line of sight of the stellar component and the relative offset between the stars and a thin dust layer.  We could assume that the stellar density follows one of a number of distributions such as a Gaussian, a log-normal, an exponential, etc.  \citet{Subramanian:2009} find that the spread in color and magnitude of red clump stars in the central region of the SMC (covering 2.5 square degrees, which includes the SMIDGE region) is best fit with a Gaussian.  Using Cepheid observations to study the 3D structure of the SMC, \citet{JD:2016a} also find a Gaussian-like distance distribution for the galaxy.  Our model also assumes a Gaussian stellar distribution along the line of sight (centered at a mean distance of $\mu$=19.07 mag) with a FWHM between 3.5-19.8 kpc which spans the range found by previous observations.  We can adapt this model to explore a variety of stellar density distributions in the future.

We apply distance and geometry offsets using the \texttt{syncmd} code in the following way:  We use the unreddened, single-distance CMD obtained as in Sec. \ref{sec:unreddenedCMD} and apply a random Gaussian distance distribution to the stars with a width $\sigma_{DM}$.  We then place a single thin dust layer at a specified position and find which stars are foreground and background to the dust, and redden the stars behind the dust as described in Section \ref{sec:addExtinction}.

In our model, we assume the dust is in a thin layer relative to the stellar distribution.  We can test this assumption with a population of stars which is distributed uniformly and independently of the dust layer.  RGB stars can serve this purpose since they are evolved stars and would be distributed regardless of the location of the dust layer.  Indeed, we do not see RGB stars covering the full range of possible extinctions, but rather we see a bimodal RGB which indicates that stars are essentially either unreddened or reddened, placing them either in front of or behind a thin dust layer. If the dust layer had a substantial thickness relative to the stellar distribution, this clear bimodality would not be evident. From stellar counts in the reddened and unreddened RGB, we conclude that the reddened fraction is $>$ 50\%.  We thus consider only dust layer positions closer to us than the centroid of the stellar distribution, as specified in Table \ref{tab:paramsvary}.

\begin{deluxetable*}{lccc}[!H]
\tabletypesize {\footnotesize} 
\tablecolumns{4}
\tablewidth{0.95\textwidth} 
\tablecaption{Model Parameters}
\tablehead{ \multicolumn{1}{l}{Parameter} & \colhead{Min} & \colhead{Max} & \colhead{Resolution} }
\startdata
 Galactic Line-of-Sight Depth$^{1}$ & 0.05 mag (3.5 kpc) & 0.275 mag (21.3 kpc) & 0.025 mag\\
 Dust-stars offset$^{2}$ & 0.02 mag (0.6 kpc) & 0.2 mag (5.7 kpc) & 0.02 mag\\
 Dust Extinction$^{3}$, log-normal median $\widetilde{A_V}$ & 0.23 mag & 0.41 mag & 0.02 mag\\
 Dust Extinction, width $\widetilde{\sigma_{A_V}}$ & 0.40 & 0.95 & 0.05 \\
 \enddata
\tablecomments{(1) The SMIDGE distance modulus is measured to be $\mu$=19.07 mag (see Figure \ref{fig:dmodtest}) indicating a distance of 65.16 kpc.  Galactic depth range in kpc is expressed as the FWHM of the Gaussian stellar distribution where $\mu$=19.07 $\pm$ 0.05 mag (lowest grid boundary for the galactic depth) corresponds to a FWHM = 3.53 kpc (1 $\sigma$ depth of 1.50 kpc) and $\mu$=19.07 $\pm$ 0.275 mag (highest grid boundary) corresponds to a FWHM = 19.81 kpc (1 $\sigma$ depth of 8.41 kpc). (2) The dust layer is offset on the near side of the stellar distribution by the distance indicated. (3) Log-normal distribution of dust extinctions where $\widetilde{A_V}$ is the log-normal median, and the dimensionless $\widetilde{\sigma_{A_V}}$ is the distribution width (see Section \ref{sec:addExtinction}.)}
 \label{tab:paramsvary}
\end{deluxetable*}

\subsubsection{Modeling the Extinction} \label{sec:addExtinction}

The \texttt{fake} routine outputs the effective temperature, surface gravity, and metallicity for each star in the CMD.  We input these into \texttt{syncmd} which uses the BEAST to compute the associated stellar spectra in order to correctly apply extinction to the model spectra. (i.e., see \citealt{2020VanDePutte, 2020Choi} for examples of using the BEAST).  We apply a \citetalias{Gordon:2003} SMC Bar $R_{V}$ = 2.74 extinction law.  The BEAST then extracts the integrated band fluxes and computes synthetic \textit{HST} photometry which we can compare to the observations.

Following \citet{Dalcanton:2015bl}, we assume the extinction has a log-normal distribution, with median $\widetilde{A_V}$ and a dimensionless width $\widetilde{\sigma_{A_V}}$.  The (normal) mean $A_V$ is related to the median $\widetilde{A_V}$ and $\widetilde{\sigma_{A_V}}$ of the log-normal by:
\begin{equation}\label{eqn:meanav}
\langle A_{V} \rangle = \widetilde{A_V}  e^{\widetilde{\sigma_{A_V}^{2}}/2}
\end{equation}
The width of the log-normal, $\langle \sigma_{A_{V}} \rangle$, can be related to $\widetilde{\sigma_{A_V}}$ by:
\begin{equation}\label{eqn:meansigav}
\langle \sigma_{A_{V}}^2 \rangle = \widetilde{A_V}^{2}  e^{\widetilde{\sigma_{A_V}^{2}}} (e^{\widetilde{\sigma_{A_V}^{2}}} - 1)
\end{equation}
or in an alternate way as in Sec. 3.1 of \citet{Dalcanton:2015bl}.  The range of extinction parameters is shown in Table \ref{tab:paramsvary}.  

It is also possible to vary the extinction curve, but since we are modeling CMDs only with the optical color $\fsf-\eof$, we are not sensitive to the difference between an SMC and a MW extinction curve (see discussion in the Appendix).  Thus we model our CMDs with the extinction law fixed to the \citetalias{Gordon:2003} SMC Bar extinction curve, which is one of the reasons we focus on this color combination.  In future work we plan to expand this technique to multiple color combinations and vary the extinction curve as well.

\subsubsection{Observational Noise Model} \label{sec:addNoise}
Accurate simulated noise is critical to our approach as it accounts for photometric errors, crowding and incompleteness through artificial star tests (ASTs) performed on the SMIDGE observations.  For details on how ASTs are used in our analysis, see Section 2.3.1 of \citet{Williams:2014} who use the same procedures for the Panchromatic Hubble Andromeda Treasury (PHAT) survey.  We employ a conservative noise model using the ``toothpick'' method of \citetalias{Gordon:2016} which is used for data that are known to be correlated between photometric bands by treating the bands independently.  The model is computed in equally spaced logarithmic flux bins and assumes that each AST entry corresponds to one artificial star.  See Sections 4.4 and 5 in \citetalias{Gordon:2016} for details on how ASTs are used to determine the noise model.

\subsection{Modeling the SMIDGE Star Formation History} \label{sec:modelingSFH} 

The SFH affects the RC/RGB stellar density, position of the unreddened RC, and morphology of CMD features.  While finding the SFH for the SMIDGE field is not the primary goal of our study, a representative SFH is key to successfully reproducing these features in the unreddened CMD, and will ensure that our model will be sensitive to potentially subtle variations in the dust extinction and geometry properties.

The RC consists of low-mass K giants in their He-burning post-RGB phase.  A K giant has a nearly constant absolute magnitude during this phase of its evolution, which places RC stars within a tight clump in the CMD.  The RC is thus less affected by age variations than the RGB in terms of CMD position.  However, both the RC and the RGB are sensitive to the metallicity and metallicity dispersion at a given age.  We model these effects simultaneously by varying both the SFR($t$) and the metallicity, Z($t$), to the ranges we specify in Table \ref{tab:sfhvary}. We assume a constant metallicity dispersion for all ages.  To determine the value of the dispersion we compare the width of the unreddened RGB in color to that of model CMDs in the following way: we define a box around the unreddened RGB in a highly reddened region where the unreddened and the reddened stars are well-separated from each other, and compare the standard deviation in color between the models and the SMIDGE data.  We find that a metallicity dispersion of 0.25 dex best reproduces the unreddened RGB width.

SFH studies exist for regions near SMIDGE (\citetalias{Weisz:2013mc}, \citealt{Rubele:2015vmc, Rubele:2018}).  Although those regions are close by, there are differences in the number counts of RC/RGB stars between their fields and SMIDGE. Adjustments to existing SFH results allow us to obtain a better fit to the observations since stellar number counts of RC and RGB stars are directly related to the SFR(t) over ages relevant for stars to evolve off the main sequence.  For example, when modeling the unreddened RC in \citetalias{YMJ:2017}, we performed a rough refinement of the SMC SFR($t$) and Z($t$) derived by \citetalias{Weisz:2013mc} to match to first order the RC morphology and stellar positions (we neglected the stellar number counts; see Sec. 3.2 in \citetalias{YMJ:2017}).  Here we need an additional SFR refinement since our method is sensitive to stellar number counts, the width of the unreddened RGB, and the morphology of the red clump.

As a starting point we use the results of \citetalias{Weisz:2013mc}, who have used deep HST photometry to study the SFH of a region in the SMC Bar near SMIDGE (shown in Figure \ref{fig:smidge_spitzer_sage}) with low internal dust extinction and an RGB stellar surface density which is somewhat lower than that of the SMIDGE region.  \citet{Rubele:2015vmc, Rubele:2018} have also examined the SFH of regions near and within SMIDGE.  Upon a SFH and age-metallicity comparison with \citetalias{Weisz:2013mc} we find that although the \citet{Rubele:2015vmc, Rubele:2018} results show higher metallicity and lower $A_V$, their results for older dominant RC ages compensate for these differences and show a consistency in the stellar number counts and morphology of the RC and the RGB.

The adjustments we made in \citetalias{YMJ:2017} to the \citetalias{Weisz:2013mc} SFR are indicated by the dashed blue line in Figure \ref{fig:sfhpanels} and include star formation enhancements at t = 500 Myr and 1.5 Gyr $\leq$ t $\leq$ 3 Gyr.  We additionally enhance the SFR in this work in order to match the total number of stars in the observed RC/RGB region ($\sim$ 12,300) where we increase the lower bound of the sampled SFR range for ages $\geq$ 3 Gyr to 1.5 $\times10^{-4} M_{\odot}$ per year.  While we explore a relatively narrow metallicity range, we find that Z($t$) offsets to lower values by 0.2 - 0.3 dex produce a better fit.  We use a set of 16 logarithmic time bins (listed in Table \ref{tab:sfhvary}) -- re-binned from the original 40 time bins of \citetalias{Weisz:2013mc} -- which range from 6.6 $\leq$ log($t$) $\leq$ 10.15.  We make this simplification since the RC/RGB populations consist of older ages and smoothing over the \citetalias{Weisz:2013mc} bins at younger ages results in a minimal variation in the relevant areas of the CMD. 

We search for a best match to SMIDGE by exploring a range around our initial estimate for SFR($t$) and Z($t$) illustrated by the width of the gray band in Figure \ref{fig:sfhpanels}.  We define coarse diagonal CMD bins representing the upper RGB, RC \& RGB, mid RGB (below the reddened RC), and lower RGB (Figure \ref{fig:syncmdregions}), and compare the stellar number counts per bin in each synthetic CMD to SMIDGE.  First we compare the unreddened, single-distance CMDs to SMIDGE, and then using forward-modeling we compare CMDs with simulated reddening and geometry as described in the next paragraph.  We search for an initial set of SFHs producing CMDs that fall within the cuts in Table \ref{tab:sfhstats} which are chosen to allow for a small variation around the SMIDGE counts.  The diagonal bins are parallel to the extinction vector which we know from the approximate slope of the RC reddening vector in \citetalias{YMJ:2017}.  They are designed to be relatively wide in color and magnitude such that reddening, geometry and noise would minimally affect the number counts.  

To account for any subtle changes to the CMD stellar number counts due to dust and geometry and due to the reddening of stars with unreddened positions bluer than $\fsf-\eof$ = 1.2 mag, we refine the initial selection of models by transforming the unreddened CMDs with \texttt{syncmd} into models incorporating geometry, dust extinction, and observational noise effects.  We use the dust extinction and geometry values we found for SMIDGE in \citetalias{YMJ:2017} (listed in the last paragraph of Sec. \ref{sec:addDistance}) as approximate parameters and follow the process in Sections \ref{sec:addExtinction} and \ref{sec:addGeometry}.  Our final SFH choice is guided by a second round of narrower cuts around the SMIDGE number counts (see Table \ref{tab:sfhstats}) which improve the fit slightly.  The resulting best-fit SFR is shown in red in Figure \ref{fig:sfhpanels}.

The SFR we find to be suitable for the SMIDGE field in Figure \ref{fig:sfhpanels} shows enhancements to the \citetalias{Weisz:2013mc} SFR at almost all ages, even those where we allowed values to vary below their result.  The most significant difference is at older ages ($\geq$ 3 Gyr) where the SFR is enhanced by a factor of 2.  The two star formation peaks at more recent times are also slightly enhanced.  These differences are expected due to the higher stellar surface brightness of the SMIDGE field compared to the \citetalias{Weisz:2013mc} field.

We note that the model CMDs overpredict the contribution of red horizontal branch (HB) stars due to known challenges in modeling the HB morphology stemming from RGB mass-loss uncertainty \citep{2009:Catelan, 2010:Gratton}.  Despite this contamination, we do not expect the HB contribution to be significant or to dominate the statistics given that we find that the HB accounts for only about 3\% of the total number of RC/RGB stars.

\begin{figure*}
  \centering 
  \includegraphics[width=1.0\textwidth]{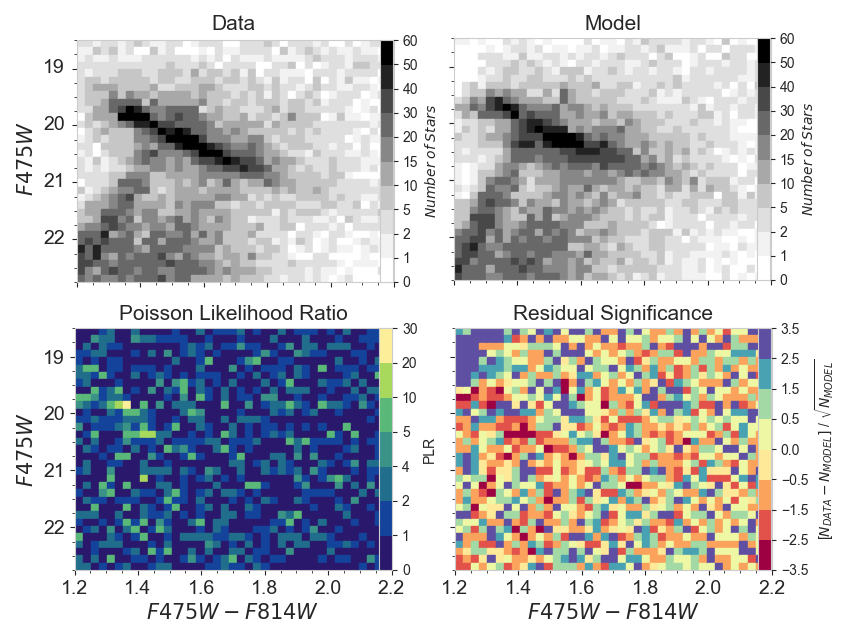}
  \caption{CMD matching of the SMIDGE color-magnitude diagram with a focus on the red clump and the red giant branch. Top: The SMIDGE observations and the best-fit model are binned in a Hess diagram to show the density of sources according to the binning scheme used in our CMD fitting. Top left: the SMIDGE data as in Figure \ref{fig:smidgeCMD}.  Top right: the best fit maximum probability model obtained as in Section \ref{sec:cmdmatching}.  Bottom left: Poisson Likelihood Ratio map shows the PLR calculated for each CMD bin using Equation \ref{eqn:lnplr}. Bottom right: Residual significance map showing the difference in the number of sources between data and model weighted by the uncertainty in the model measurement.  The best-fit model in the upper right has the following parameters: dust-stars offset of 0.10 mag, 1$\sigma$ distance modulus spread of 0.15 mag, dust extinction with a log-normal median $A_V$ = 0.33 mag, and width $\sigma_{A_V}$ = 0.65.}
 \label{fig:cmdpanel}
\end{figure*}

\subsection{Comparing Model and Observed CMDs} \label{sec:pdfanalysis}

We generate a grid of model CMDs for each dust and geometry parameter combination from Table \ref{tab:paramsvary} for a total of 12,000 models.  To compare each model CMD to the SMIDGE observations, we calculate the probability that a model produces the observations by comparing the number of stars in the respective CMDs in a way which is sensitive to CMD features (described below).  This calculation results in the probability density as a function of each dust extinction and geometry parameter, or a combinations of parameters.  The probability density function (PDF) can in turn inform us about the confidence interval for each parameter based on a set probability threshold.

To compare the number of stars in the model and data CMDs, we bin each model CMD in an identical way to the observed CMD and calculate the probability, from a Poisson distribution, that a model with a predicted number of stars $n_i$ produces an observed $m_i$ stars in bin $i$:

\begin{equation}\label{eqn:poissonprob}
    P_{i} = \frac {m_i ^ {n_i} } {e^{m_i} n_{i}!}
\end{equation}

We follow the recommendation of \citet{Dolphin:2002, 2013:Dolphin} who suggest that in the case of binned CMD data, the Poisson probability distribution should be used instead of the canonical Gaussian distribution, e.g. as in a $\chi^{2}$ minimization \citep{Mighell:1999}. The results for all CMD bins are multiplied to obtain the total probability for each model-data comparison.

The probability can be measured for any binning of the CMD in color and magnitude.  There are several considerations to take into account when defining the binning.  First, if there are features in the CMD which we would like to be sensitive to, we must choose binning sufficiently small such that these features are preserved.  For our purposes, we want to be sensitive to features such as the unreddened and the reddened RC, the tip of the RGB, and the unreddened RGB, among others.  The smallest (narrowest) of these features are the unreddened RGB (which spans $\sim$ 0.1 mag in color) and the unreddened RC (spanning $\sim$ 0.5 mag in magnitude).  Second, we also make sure that our bins are not too small that we end up with a significant number of bins ($\geq$ 5 \%) containing no sources.  Finally, we must choose either uniform binning, where all CMD bins are of equal rectangular size, or irregular binning, to fit parts of the CMD in a bin of a special size if we have a sense for known errors or uncertainties in the data.  Here we follow \citetalias{Dolphin:2002} in choosing uniform binning. Our final bin size is 0.025 mag in color and 0.125 mag in magnitude, which finely samples the CMD features of interest while minimizing the number of empty bins.  With our SFH analysis in Section \ref{sec:modelingSFH} we ensure that the difference in the total number of stars in the RC/RGB region between the SMIDGE CMD and each model CMD is insignificant.  The variation in this number due to noise in each model CMD is addressed in Section \ref{sec:uncertainty}.

The model with the highest probability is our best-fit model. To visualize the goodness of fit per bin, it is convenient to calculate a maximum likelihood statistic based on this probability.  Similarly to \citetalias{Dolphin:2002} who applied this statistic to find the best star formation history for nearby galaxies, we calculate the Poisson likelihood ratio (PLR) for each model-data comparison.  Many studies use the PLR for CMD fitting, primarily to analyze SFHs \citep{Dolphin:1997, Aparicio:1997, Dolphin:2002, HarrisZaritsky:2009,2011:Weisz}.  The PLR is defined as the ratio of the probability of drawing $n_i$ stars from model $m_i$ to the probability of drawing $n_i$ stars from model $n_i$ and has the following form:

\begin{equation}\label{eqn:plr}
    PLR_i = \frac{m_i^{n_i} e^{n_i}} {n_i^{n_i} e^{m_i}}
\end{equation}

We simulate distance, dust extinction and observational noise simultaneously on our unreddened synthetic CMDs and then measure the degree of match.  In this sense our synthetic model CMD is stochastically sampled, and to calculate the likelihood -- $L(n_1, n_2)$ -- that two samples, $n_{1}$ and $n_{2}$, come from the same underlying model, $m$, we use:

\begin{equation}\label{eqn:pdistr} 
L(n_{1}, n_{2}) = \int_{0}^{\infty} \frac {m^{n_1}} {e^m n_1!} \frac {m^{n_2}} {e^m n_2!} dm = \frac {0.5^{(n_1+n_2+1)} (n_1+n_2)!} {n_1! n_2!}
\end{equation}

The equivalent PLR for this expression when $n_1$ is the observed data and $n_2$ is the model is (see Sec. 2.3 of \citetalias{Dolphin:2002} for details):

\begin{equation}\label{eqn:lnplr} 
-2 \ln PLR = -2 \ln \frac{L(n_{1}, n_{2})} {L(n_{1}, n_{1})}
\end{equation}

We sum the log of the PLR for all bins to obtain a total PLR for each model-data comparison.

The best-fit maximum probability model and the corresponding PLR results per bin are shown in Figure \ref{fig:cmdpanel}.  The residual significance in the lower right panel measures the fit quality by subtracting the model from the observed CMD and dividing this difference by the uncertainty in the model count, $(n-m) / \sqrt{m}$.  This result is simply the residual in units of standard deviation (e.g., a value of $+1$ or $-1$ is a 1$\sigma$ outlier).  

There is a very good agreement between the overall number of stars in the RC/RGB CMD, but some weaknesses in the match are evident.  The overestimated contribution of the horizontal branch (HB) discussed in Section \ref{sec:modelingSFH} can be seen just below the red clump in both the CMD data-to-model comparison (as a slight bulge), and in the PLR and residual significance maps (as a green/yellow and a red patch, respectively).  One consequence of this HB overestimate is the appearance of a ``dip'' in the reddened model RC (near the location of the unreddened RGB) even though the measured RC slopes of the data and the best fit model are nearly identical.  The steeper model RGB slope is another potential consequence of this contribution.  Both the PLR and the residuals maps show a poor match (an overestimation) in the number of (reddened and unreddened) RC stars which may have to do with the overestimation of the number of stars in the HB.  In the Appendix we show examples of models which are bad fits to the data.

\begin{figure*} 
  \centering
    \includegraphics[width=1\textwidth]{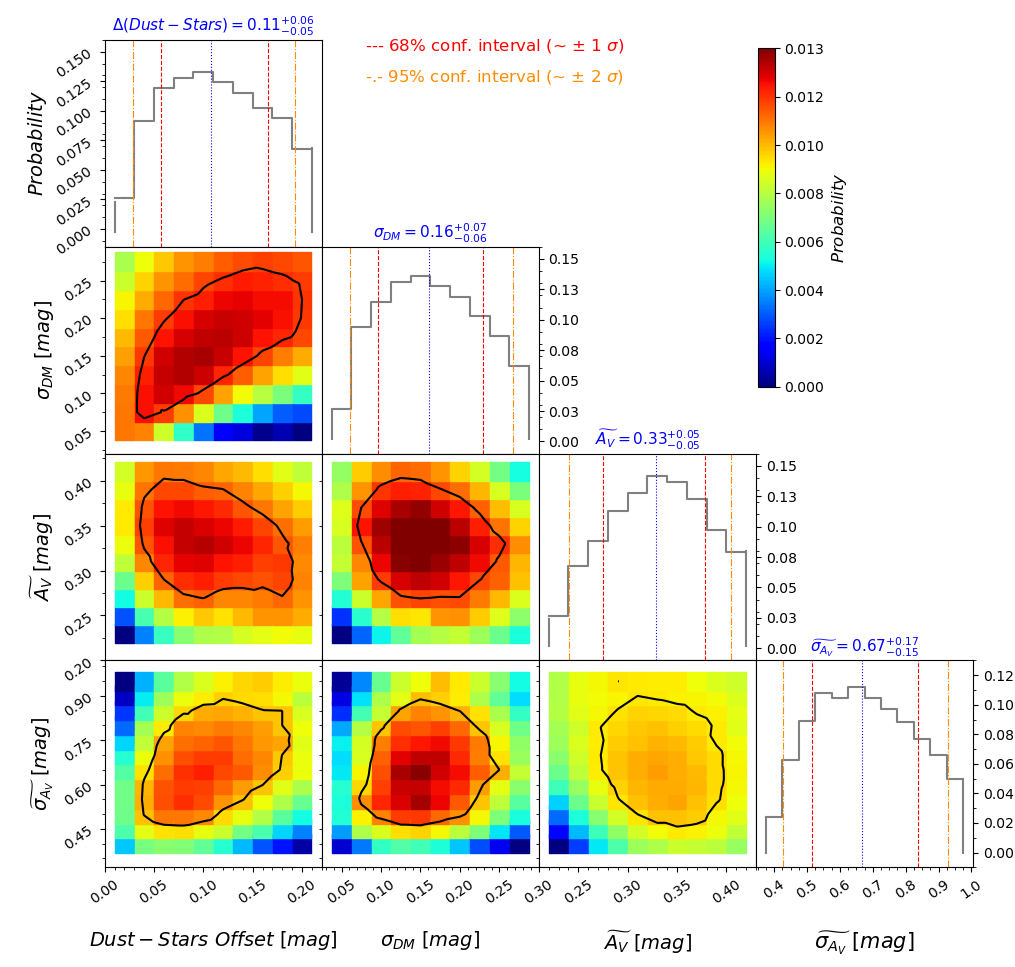}
  \caption{1D and 2D probability distribution functions based on CMD matching results for the 3D geometry and dust extinction of the SMIDGE field. The 1D 68\% and 95\% confidence intervals fall between the red dashed and orange dot-dashed lines, respectively. The 50\% probability threshold shown in dotted blue indicates the most likely value of the corresponding parameter. If the PDF followed a normal distribution, then the 68\% confidence interval would be equivalent to $\pm 1 \sigma$ uncertainty and the 50\% blue line would indicate the mean.  2D probability distribution functions show the correlation among the geometry and extinction parameters.  The black contours represent the 2D 68\% confidence interval threshold.}
  \label{fig:pdf}
\end{figure*}

\begin{deluxetable*}{ll}[!H]
\tablecolumns{3}
\tablewidth{0.8\textwidth}
\tablecaption{SMIDGE 3D Structure and Dust Extinction Results}
\tablehead{ \multicolumn{1}{l} {Parameter} & \multicolumn{1}{l}{PDF 50\% median and 68\% confidence intervals}}
\startdata
 Dust-stars offset
 & 0.11 $^{+0.06}_{-0.05}$ mag; {\hspace*{1.4cm}} 3.22 $^{+1.69}_{-1.44}$ kpc \\[0.15cm]
 Distance modulus spread, $\sigma_{DM}$
 & 0.16 $^{+0.07}_{-0.06}$ mag; {\hspace*{1.4cm}} FWHM $\approx$ 11.3 kpc \\[0.15cm]
 Reddened fraction, $f_{RED}$ 
 & 0.73 $^{+0.13}_{-0.13}$ \\[0.15cm]
 Extinction Log-normal, $A_{V}$
 & $\widetilde{A_V}$ = 0.33 $^{+0.05}_{-0.05}$ mag; {\hspace*{0.8cm}} mean $\langle A_V \rangle$ = 0.41 $\pm$ 0.09 mag \\[0.15cm]
 Extinction Log-normal width, $\sigma_{A_V}$
 & $\widetilde{\sigma_{A_V}}$ = 0.67 $^{+0.17}_{-0.15}$ {\hspace*{1.4cm}} mean $\langle \sigma_{A_V} \rangle$ = 0.097 $\pm$ 0.003\\
\enddata
\tablecomments{The dust layer is on the near side of the stars. The log-normal median $\widetilde{A_V}$ and $\widetilde{\sigma_{A_V}}$ are converted into the mean (normally-distributed) $\langle A_V \rangle$ and $\langle \sigma_{A_V} \rangle$ using Equations. \ref{eqn:meanav}-\ref{eqn:meansigav}.}
 \label{tab:results}
\end{deluxetable*}

\subsection{The Probability Density Function and Uncertainties in the Results} \label{sec:uncertainty}

Each model CMD is subject to two main sources of uncertainty.  One is a result of number statistics due to the small number of stars in each CMD bin.  Additionally, each CMD contains an intrinsic uncertainty in the stellar positions due to the random photon noise and crowding effects (simulated by the BEAST).  These effects result in an inherent variability in the model predictions when a model is generated multiple times with the same set of parameters.

We generate synthetic CMDs over a well-sampled evenly-spaced grid of models based on the ranges of geometry and dust extinction parameters in Table \ref{tab:paramsvary}. To account for the sources of uncertainty above, we generate each CMD in the grid 50 times and calculate the average number of stars per CMD bin.  Each model is fit to the data by the process in Section \ref{sec:pdfanalysis}.  The repeated trials average out the noise.

\begin{figure*} 
  \centering
    \includegraphics[width=1\textwidth]{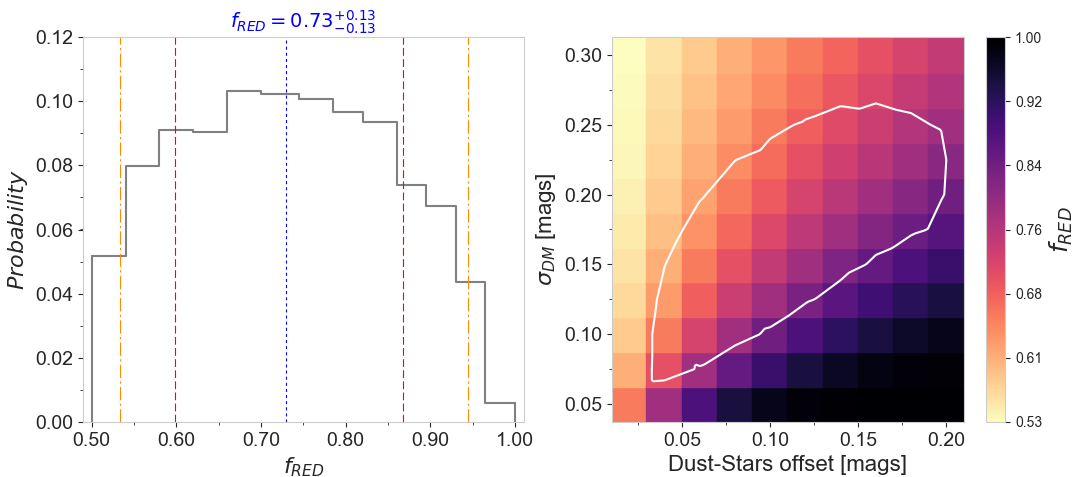}
  \caption{The result of the SMC 3-dimensional structure on the reddened fraction, $f_{RED}$, in terms of the relationship between the spread in the stellar distribution and the location of the dust layer relative to the stars. As expected, $f_{RED}$ increases as the dust is farther removed from the stars (on the near side) while the stellar distribution remains compact.  Lines of constant reddened fractions align with the white 68\% confidence interval contours of the dust-stars offset and $\sigma_{DM}$ 2D PDF in Figure \ref{fig:pdf}.  The observations indicate $f_{RED}$=0.73$^{+0.13}_{-0.13}$. (See Section \ref{sec:results}.)}
  \label{fig:fred}
\end{figure*}

In addition to finding the best-fit model (see end of Section \ref{sec:pdfanalysis}), we also calculate the median and the confidence intervals of the 1-dimensional PDFs of each parameter.  We define the 50$^{th}$ percentile of the cumulative probability as the PDF median, and the 16-84$^{th}$ percentile confidence interval threshold as a measure of the PDF width (68$\%$ around the median), or the uncertainty in our results, informing us of the range of parameter values which contains the corresponding proportion of the probability.  If the probability distribution were Gaussian, then the 68\% confidence interval would correspond to $\pm$ one standard deviation, $\sigma$, and the 50$^{th}$ percentile would correspond to the mean.  The results of this analysis are shown in Figures \ref{fig:pdf} and \ref{fig:fred}.

To obtain a 1D PDF for each parameter, we marginalize over all other parameters after one step of renormalizing.  In this step, we assume that the probability of a model outside the grid is $\sim$ 0, thus we cover a total probability of 1 in the grid, where each model is normalized accordingly.  We marginalize over various dimensions of the grid to produce 1-D and 2-D PDFs for each parameter and parameter pair.  In this way we can find desired confidence intervals for our parameters. 

\section{Results} 
\label{sec:results}
The results for the 3D geometry and dust extinction parameters and their 68\% confidence interval thresholds are presented in Table \ref{tab:results} and Figures \ref{fig:pdf} and \ref{fig:fred}.  We find that the dust layer is offset on the near side of the centroid of the stellar distribution by 3.22 $^{+1.69}_{-1.44}$ kpc, and that the CMD is best fit with a stellar distribution along the line of sight with a 1 $\sigma$ width of 0.16 $^{+0.07}_{-0.06}$ mag, or an equivalent FWHM of 11.3 kpc.  The reddened fraction of stars is $f_{RED}$=0.73$^{+0.13}_{-0.13}$ with the dust on the near side of the stellar midplane.  The log-normal of dust extinction has a median of $\widetilde{A_V}$=0.33 $\pm$ 0.05 mag and width $\widetilde{\sigma_{A_V}}$=0.67 $^{+0.17}_{-0.15}$.  This corresponds to a mean $\langle A_V \rangle$ = 0.41 mag and a width $\sigma_{A_V}$ = 0.097 (where the foreground $A_V$ = 0.18 mag applied to model the SFH is not a part of this result, thus it is solely attributable to SMC dust.).

There is a lack of strong correlation between the geometry and the extinction parameters in the fitting procedure.  In particular, $\widetilde{A_V}$ and $\widetilde{\sigma_{A_V}}$ do not appear correlated with the dust-stars offset or the line-of-sight depth.  The interpretation of this result is that since stars lie either in front of or behind the thin dust layer, they experience either all of the dust column or none of it (as discussed in Section \ref{sec:addExtinction}).  The geometrical arrangement of the stars and the dust is therefore independently constrained from the extinction properties of the dust layer.

\begin{figure*} 
  \centering
    \includegraphics[width=1.\textwidth]{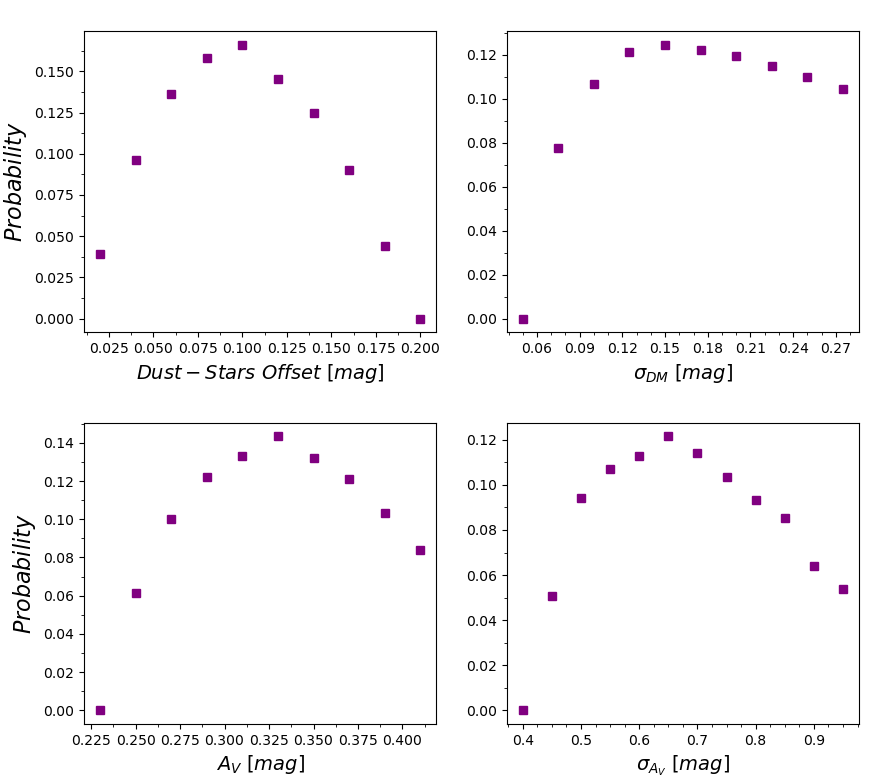}  
   \caption{Model grid slices in multidimensional space showing the relationship between the best-fit CMD model parameters. Each point represents the relative probability for a model corresponding to a parameter value shown on the x-axis where the rest of the parameters for this model are fixed at their best-fit values.}
   \label{fig:gridslice}
\end{figure*}

Figure \ref{fig:pdf} and Figure \ref{fig:gridslice} show that all parameters are well-constrained.  Figure \ref{fig:gridslice} shows a slice through the multidimensional grid, with all parameters held fixed at their best-fit values from Section \ref{sec:pdfanalysis}, aside from the value on the x-axis.  Both figures show that small values of the line-of-sight depth, $\sigma_{DM}$, are strongly ruled out.

One of the most strongly constrained parameters from the observations is $f_{RED}$, which can be found simply from counts of RGB stars in the unreddened and reddened branches. $f_{RED}$ is a purely geometrical effect which results from the relative arrangement of the dust-stars offset and the stellar distribution along the line-of-sight ($\sigma_{DM}$).  The correlation between the line-of-sight depth and the dust-stars offset is the strongest among all the parameters.   $f_{RED}$ can be predicted theoretically from the relationship between $\sigma_{DM}$ and the dust-stars offset as illustrated in the right-hand panel of Figure \ref{fig:fred}.  Regions of constant $f_{RED}$ indicate that the closer the dust layer is to the centroid of the stellar distribution, the smaller the line-of-sight depth required to preserve the same reddened fraction.  Indeed, the 68\% confidence intervals of the 2D PDF of the dust-stars offset -- $\sigma_{DM}$ trace lines of constant $f_{RED}$. Our results for $f_{RED}$ = 0.73$^{+0.13}_{-0.13}$ are consistent with the 68\% confidence interval.

\section{Discussion}
\label{sec:discussion}

Our two main findings about the SMC galactic structure are that the SW bar of the galaxy shows a significant spread in the distribution of stars along the line of sight of more than 10kpc FWHM, and also that the dust layer, which is clearly observed from its extinction effects, is offset from the centroid of the stars on the near side of the SMC.  We obtained a similar result for the line-of-sight depth (10 $\pm$ 2 kpc) in \citetalias{YMJ:2017} employing a simple model for the reddened RC which helped explain the observed offset in the extinction curve in the SMC and the LMC.  

Our line-of-sight depth finding is consistent with a number of other studies of the SMC structure employing various methods which generally suggest that the SMC is significantly elongated along the line of sight.  Depending on the region studied and the method used, results vary and indicate galactic depths of less than 10 kpc and up to $\sim$ 23 kpc in the outer regions of the galaxy \citep{HatzidimHawkins:1989, Gardiner:1991, HarrisZaritsky:2004, Subramanian:2009, Subramanian:2012, Nidever:2013, Rubele:2018, Muraveva:2018}.  

Distance tracers show a number of features in the SMC, such as a distance gradient, galactic depth, and a distance bimodality.  The distance gradient observed generally suggests that the eastern regions are closer than the western ones \citep{JD:2016a, Subramanian:2012, Ripepi:2016, Subramanian:2017, Muraveva:2018}.  The galactic depth has been observed by finding the distribution of RR Lyrae and Cepheid variables \citep{HarrisZaritsky:2004, Kapakos:2012, Haschke:2012smc, JD:2016a, Muraveva:2018} and the luminosity dispersion of red clump stars \citep{Gardiner:1991, Nidever:2013, Subramanian:2017}.  Using the latter also shows a distance bimodality \citep{Subramanian:2017} with two distinct bodies of stellar structures in the eastern regions of the galaxy.  Although we do not study the distance gradient here, we note that the distance modulus we find via CMD matching discussed in Section \ref{sec:modelingSFH} of $\mu$=19.07 mag indicates that the SMIDGE region in the SMC's SW Bar is offset from the center of the SMC (found to be at a distance modulus of $\mu$=18.96 mag \citep{Scowcroft:2016hm}), consistent with the studies above in that it indicates the western region is farther away.

A significant line-of-sight depth is also consistent with an LMC/SMC interaction.  Numerical models for the SMC and the LMC indicate that the SMC geometric features, as well as the internal kinematics of the Magellanic Clouds, can be explained by repeated interactions between the two galaxies \citep{Besla:2012, Besla:2016}.  The models of \citet{Besla:2012} show that ram pressure decreases the velocity of the SMC's ISM and plays a role in modifying the properties of the ISM distribution, such as its location and velocity.  Our observations that the dust layer is thin and results that the dust is positioned on the near side of the SMC relative to the stellar centroid (therefore placing the dust nearer the LMC as well) may be an expected consequence of such a ram pressure effect.  \citet{Choi2018} measured the dust reddening and 3D structure of the LMC using RC stars and discovered a new stellar warp toward the SMC in the outer disk (and an LMC line-of-sight depth of $\sim$7 kpc).  It is clear from our study and others that the line-of-sight geometry of both Magellanic Clouds is complicated, due to their interactions.

We compare our CMD fitting result for the average SMIDGE dust extinction with $A_V$ based on dust mass surface densities, $\Sigma_{M_d}$, derived from IR emission observations by the HERschel Inventory of The Agents of Galaxy Evolution \citep{2014:Gordon}.  We fit the IR dust SED with a modified black body dust emission model as in  \citet{2018:Chiang} to extract the average dust mass surface density in the SMIDGE footprint, and find $\Sigma_{M_d}$ = 1 $\times$ 10$^5$ M$_{\odot}$ kpc$^{-2}$.  Using $A_V/\Sigma_{M_d}$ = 0.7394 $mag/10^5M_{\odot}$ kpc$^{-2}$ from the Draine \& Li model and from \citet{2014:Draine}, the IR-derived extinction is $A_V$ $\approx$ 0.75.  This overprediction of the observed extinction -- here by a factor of $\sim$ 1.8 -- is consistent with other comparisons between IR-derived optical extinction results using the Draine \& Li dust model and alternative extinction methods.  \citet{Dalcanton:2015bl} who use CMD fitting of RGB stars to derive the distribution of dust in M31 find that the dust model $A_V$ estimates are too high by a factor of $\sim$ 2.5 when comparing their results to those of \citet{2014:Draine}.  Similarly, \citet{2016:Planck} show a discrepancy of a factor of 1.9 in $A_V$ estimates from all-sky Milky Way optical photometry of quasi-stellar objects.  This discrepancy can be attributed to the Draine \& Li model overestimating the dust mass from IR emission, which would imply that dust grains are more emissive than the model assumes.  However, it is notable that M31 and the Milky Way are both at relatively high metallicity.  Our comparison is among the first of its kind for a low-metallicity environment containing dust which potentially has very different properties.

Further, an SMC $A_{V}$ comparison can be made with the work of \cite{Hagen:2017} who use UV, IR and optical observations integrated into 200$''$ regions to model the SMC SED and map $A_{V}$ for the galaxy.  Half of their SMC regions have a mean $\langle A_V \rangle <$ 0.25, with results for the SMIDGE region varying between $A_{V}$ = 0.2 - 0.75 mag.  Our results of mean $\langle A_V \rangle$ = 0.41 are on the high end which is expected since the SMIDGE field was specifically picked as one of the dustiest spots in the SMC.  Our upcoming study will resolve the SMIDGE field into smaller regions which will allow for a better comparison with \cite{Hagen:2017}.

\citet{Dalcanton:2015bl} studied the distribution of dust in M31 by modeling the NIR CMD of red giant branch stars to obtain results for the same set of dust and geometry parameters as the ones we explore here -- $\widetilde{A_V}$, $\widetilde{\sigma_{A_V}}$, and $f_{RED}$.  Although they approach modeling the RGB and RC quite differently for a number of reasons, they also observe an RGB broadening and a bimodality due to the combined effects of dust extinction and geometry, and rely on modeling these observed effects to form the basis of their conclusions.  Their study explores regions in the star-forming disk of M31 which generally have a more widely-varying and a higher $A_{V}$ than SMIDGE.  They do not observe a correlation between $A_{V}$ and $f_{RED}$ at high extinctions (median $A_{V}\gtrsim$ 1 mag).  A correlation between the two at lower extinctions is interpreted not as a physically-driven result, but rather as a result expected from their fitting approach (which relates $f_{RED}$ to $A_{V}$ through a "filling factor" describing the areas of the gas cloud and the analyzed pixel such that an increasingly smaller filling factor for dusty gas is due to a decreasing median extinction and also contributes to a decreasing geometric reddened fraction).  At the lowest extinctions they observe, which are comparable to the extinctions we find in SMIDGE, (median $A_{V}\lesssim$ 0.3 mag), the broadening of the RGB in their NIR CMDs is not significant, thus their do not have a strong constraint on $f_{RED}$.

The LMC, also known to have an appreciable depth along the line of sight \citep{Subramanian:2009} and a relatively low metallicity at 1/2 solar \citep{RussellDopita:1992}, is another suitable target for this RC/RGB dust and geometry modeling technique.  Our goal is to apply this same approach to data from the LMC METAL survey $HST$ \citep{2019:Roman-Duval} to study dust extinction and geometry properties probing the LMC extinction curve and extinction distribution, and 3D structure.

\subsection{Calculating \texorpdfstring{$A_V/N_H$}{}}
\label{sec:dust-to-gas}

We measure the average H column density in the SMIDGE field using HI observations from \citet{1999Stanimirovic} using the ATCA and Parkes telescopes and $^{12}$CO J=(2$-$1) observations of the SW Bar with the APEX telescope (Rubio et al. in prep).  We extract the average integrated intensities of the HI and CO lines over the full coverage of the ACS imaging from SMIDGE.  We convert the HI line integrated intensity to column density assuming no opacity correction.  For H$_2$, we must adopt a CO-to-H$_2$ conversion factor.  The behavior of the conversion factor as a function of metallicity is the subject of much investigation \citep{2013Bolatto}.  For the SMC, X$_{\rm CO}$ has been found to be larger than the MW value of X$_{\rm CO} = 2\times 10^{20}$ cm$^{-2}$ (K km s$^{-1}$) by a variety of studies \citep{2011Leroy, 2014:Roman-Duval}.  The SMIDGE field is atomic gas dominated, so for a wide range of conversion factors N$_H$ does not change dramatically depending on the conversion factor choice.  We convert the CO (2-1) line to (1-0) using a $R_{21} = (2-1)/(1-0) = 0.7$ and we use X$_{CO}$ values between the MW and 10 times the MW value to obtain N$_{H2}$, which we multiply by 2 to obtain N$_{H}$.  We calculate a mean $\langle A_V \rangle$ = 0.41 $\pm$ 0.09 mag.  The final range of N$_{H}$ for SMIDGE is $9.7\times 10^{21}$ to $1.3\times 10^{22}$ cm$^{-2}$.  This yields $\langle A_V \rangle/N_H = 3.2-4.2 \times 10^{-23}$ mag cm$^2$ H$^{-1}$.  These values are substantially lower than the canonical MW $\langle A_V \rangle/N_H = 5.3\times 10^{-22}$ \citep{1978Bohlin, 2009Rachford} by more than an order of magnitude. 
\citetalias{Gordon:2003} measure the $HI$ column density from the UV Ly$\alpha$ absorption profile for four sightlines in the SMC Bar and find $A_V/N_H = 7.7 \times 10^{-23}$ mag cm$^{-2}$ H$^{-1}$, which is only a factor of $\sim$ 2 higher than our result.  \citet{2014:Roman-Duval} measure a comparable ratio in the SMC using Herschel IR observations where they find the gas-to-dust ratio to be between 4 and $\sim$ 10 times the MW value in the diffuse and dense ISM, respectively, depending on the methodology used.

\section{Conclusions and Further Work}
\label{sec:conclusions}
We model the dust extinction and 3D geometry properties of the SMIDGE field in the SW Bar of the SMC using the optical $\fsf-\eof$ CMD of RC and RGB stars.  We find the following:
\begin{enumerate}[1.]
\item The distance modulus for the midpoint of the stellar distribution within the SMIDGE region is 65.2 kpc ($\mu$ = 19.07 mag).
\item The CMD is best fit with a 1$\sigma$ line-of-sight depth $\sigma_{DM}$ = 0.16 $^{+0.07}_{-0.06}$ mag, or an equivalent FWHM of 11.3 kpc. 
\item The dust layer is offset on the near side of the stars by 3.22 $^{+1.69}_{-1.44}$ kpc and is located at a distance modulus of $\mu$ = 18.96 mag (61.94 kpc).
\item The combination of dust position and stellar distribution results in a 73$^{+0.13}_{-0.13}$ \% reddened fraction of stars.
\item The distribution of dust extinction has a mean $\langle A_V \rangle$ = 0.41 $\pm$ 0.09 mag and width $\langle \sigma_{A_V} \rangle$ = 0.097 $\pm$ 0.003 mag (log-normal median $\widetilde{A_V}$ = 0.33$^{+0.05}_{-0.05}$ mag and dimensionless width $\widetilde{\sigma_{A_V}}$ = 0.67 $^{+0.17}_{-0.15}$).
\end{enumerate}

The SMC's dust extinction curve is notably different from that of the Milky Way in terms of the lack of a strong 2175 $\angstrom$ bump and a steep rise in the UV \citepalias{Gordon:2003}.  Here we base our model grid on the \citetalias{Gordon:2003} SMC Bar extinction law ($R_{V}$=2.74) as discussed in Section \ref{sec:addExtinction} since our 2017 work \citepalias{YMJ:2017} showed consistency with the \citetalias{Gordon:2003} results.  To probe different extinction laws, including a potential extinction law mixture of SMC and Milky Way extinction curves (see Appendix), in subsequent work we will use multiple color combinations for a similar analysis, fixing the geometric parameters of the SMC at the values found here.  

\acknowledgements
We thank the anonymous referee for a very thorough and helpful report that substantially improved the paper. This work was supported in part by NASA through grant number HST-GO-13659 from the Space Telescope Science Institute, which is operated by AURA, Inc., under NASA contract NAS5-26555. D.R.W. acknowledges support from an Alfred P. Sloan Fellowship and an Alexander von Humboldt Foundation Fellowship. These observations are associated with program \# GO-13659 and are based on observations made with the NASA/ESA Hubble Space Telescope. We made extensive use of NASA’s Astrophysics Data System bibliographic services. This research made use of Astropy, a community-developed core Python package for astronomy (Astropy Collaboration et al. 2013), NumPy (Van Der Walt et al. 2011), Matplotlib (Hunter 2007), the \texttt{syncmd} tool for BEAST-based post-processing of Synthetic Stellar Populations (C. Johnson), the BEAST (Bayesian Extinction and Stellar Tool; K. Gordon), and the \texttt{MATCH} CMD fitting software package (A. Dolphin).

\section*{Appendix}
\label{sec:appendixA}

We perform tests to determine the sensitivity of our technique to a number physical- and technique-specific factors in our attempt to measure quantities of interest.

Our sensitivity to the extinction law is limited since extinction curve differences in the optical part of the spectrum are small, and our analysis is based only on optical CMDs.  We test how sensitive we are to differentiating between an SMC and a MW extinction curve by simulating CMDs with a mixture of an SMC-like \citetalias{Gordon:2003} $R_{V}$=2.74 law and a Milky Way-like \cite{Fitzpatrick:1999} $R_{V}$=3.1 law.  We do this by varying $f_{A}$, which is the fraction corresponding to the mixture between the two extinction laws (see Sec. 3.2 of \citetalias{Gordon:2016}). Although we see a trend of increasing PLR (worse match) as the model moves away from an SMC-like extinction, a pure Milky Way-like law is within the SMC noise model uncertainties.  Thus we are not sensitive to the extinction law to a sufficient degree using the optical CMD only.

\begin{figure} 
  \centering 
    \includegraphics[width=0.45\textwidth]{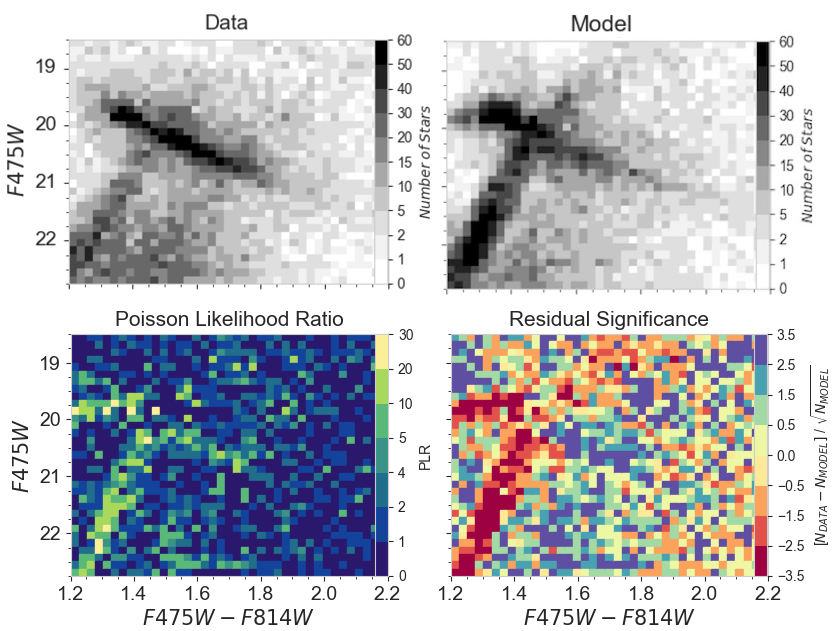}
    \includegraphics[width=0.45\textwidth]{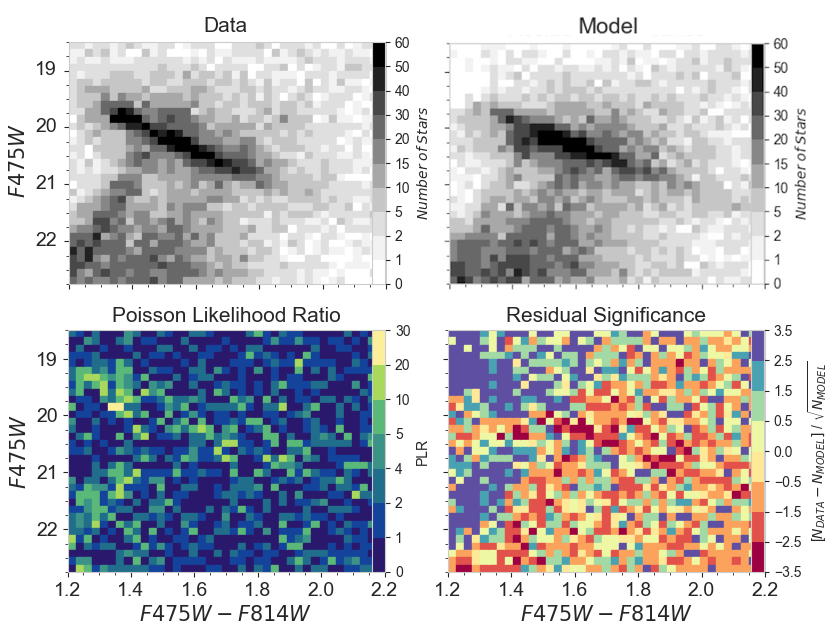}
    \includegraphics[width=0.45\textwidth]{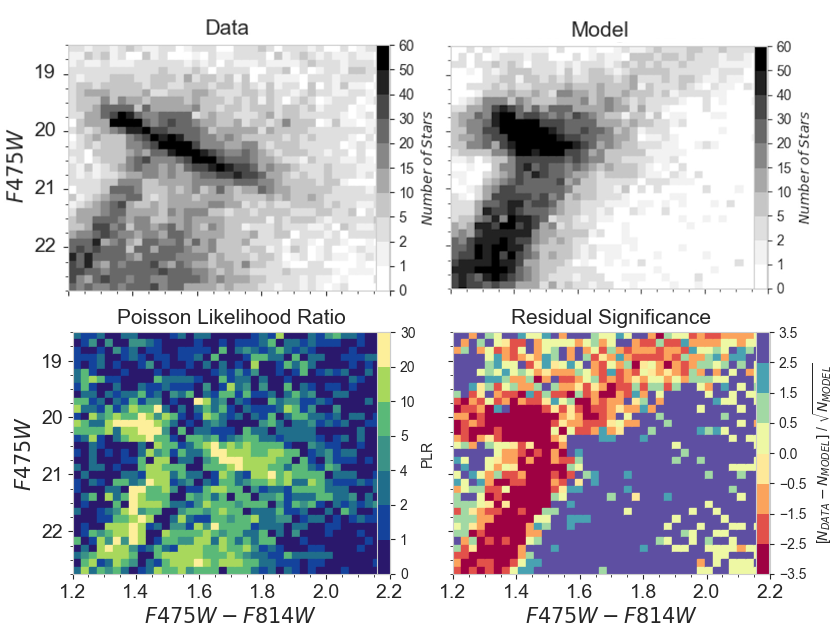}
   \caption{Examples of bad model fits where all but one parameters are at their best fit value listed in Section \ref{sec:results}. Top panel: Dust layer is aligned with the center of the stellar distribution causing a 50\% reddened fraction. Mid-panel: The spread of the stellar distribution is 1/2 of its best-fit value ($\sigma_{DM}$=2.25 kpc which results in a distribution with a FWHM = 5.3 kpc. Bottom panel: The mean of the log-normal distribution of dust extinctions is $A_{V}$=0.1 mag. )}
   \label{fig:badmodels}
\end{figure}

While Figure \ref{fig:cmdpanel} shows our best fit model, here we show examples of bad model fits where all parameters are held fixed at their minimum-PLR values, except for one parameter which is fixed at a bad-fit value.  The PLR and the residual significance plots show the mismatch clearly.  The top panel of Figure \ref{fig:badmodels} shows the result of placing the dust layer in the middle of the stellar distribution causing a $\sim$ 50\% reddened fraction.  We can see that both the RC and the RGB have an overdensitiy of stars which we do not see in the data.  

The middle panel shows a model which underestimates the spread of the stars along the line of sight (FWHM = 5.3 kpc while our results point to a FWHM $\approx$ 11.3 kpc).  We clearly see that the model lacks a sufficient number of unreddened stars extending to brighter magnitudes (which would result from a higher line-of-sight depth).  Correspondingly, the model overestimates the number of reddened stars (calculated to be $f_{RED}$) since the smaller spread in the stellar distribution places a higher fraction of stars behind the dust (see Fig. \ref{fig:fred}).

It can be also seen that the slope of the reddened portion of the red clump is incorrect since a smaller galactic depth produces a shallower RC slope, and in our case a mismatch to the data.  Indeed, the main measurement of our previous work \citepalias{YMJ:2017}, which used the slope of the reddened RC across multiple-color CMD combinations to measure the shape of the SMIDGE extinction curve, showed an offset from previous extinction curve measurements.  This result was due to the measured steepness of the RC reddening vector which resulted in higher extinction values across all CMD combinations.  Using a toy RC model to simulate an SMC depth, we showed that the offset was consistent with a stellar distribution with a FWHM = 10 $\pm$ 2 kpc along the line of sight.

The bottom panel of Figure \ref{fig:badmodels} shows a model which underestimates the amount of dust by incorporating a distribution of dust extinctions with $\widetilde{A_V}$=0.1 (contrasted with our result of $\widetilde{A_V}$=0.3).  This model also results in a reddened fraction $f_{RED}$=0.7, since the extinction has no bearing on the dust-stars geometry (and vice-versa) as discussed in Section \ref{sec:discussion}.  However, since $\widetilde{A_V}$ changes both the color and magnitude of stars, we can see the effect in both CMD dimensions for the RC and the RGB as they tend to be closer to their unreddened positions (see the left panel of Figure \ref{fig:syncmdregions}).

\bibliographystyle{yahapj}
\bibliography{dustbib}
\end{document}